# Modality contribution score: A per-patient framework for quantifying the relative diagnostic contribution of structural MRI and amyloid PET in Alzheimer's disease

Dawa Chyophel Lepcha [a, b, c], Aaliya Ali [c, d], Sophie A. Martin [e, f], Deepika Koundal [g, h], Pierrick Coupé [i], Shabbir Syed Abdul [a, b, j, *]

[a] *Graduate Institute of Biomedical Informatics, College of Medical Science and Technology, Taipei Medical University, Taipei, 235, Taiwan.*
[b] *International Center for Health Information Technology, College of Medical Science and Technology, Taipei Medical University, Taipei, 235, Taiwan.*
[c] *Biomedical Sensors & Systems Lab, University of Memphis, Memphis, TN 38152, USA*
[d] *Centre of Research Impact and Outcome, Chitkara University, Rajpura- 140417, Punjab, India*
[e] *UCL Hawkes Institute, University College London, London, UK*
[f] *Queen Square Institute of Neurology, University College London, London,UK*
[g] *School of Computer Science, UPES, Dehradun, 24800, Uttarakhand, India*
[h] *University of Eastern Finland, FI-70210 Kuopio, Finland*
[i] *CNRS, University Bordeaux, Bordeaux INP, LABRI, UMR5800, Talence F-33405, France*
[j] *School of Gerontology and Long-term Care, College of Nursing, Taipei Medical University, Taipei, Taiwan, 110.*

**Abstract**

Multimodal neuroimaging combining structural MRI and positron emission tomography (PET) captures complementary structure-function relationships across the Alzheimer's disease (AD) continuum, yet existing artificial intelligence systems produce a single diagnostic label without quantifying which imaging modality drove that decision for a specific patient. We introduce the Modality Contribution Network (MCNet) and the Modality Contribution Score (MCS), the first per-patient attribution framework quantifying the shift in modality dominance from structural atrophy to amyloid and metabolic dysfunction across the cognitively normal to MCI to AD continuum. MCS is normalised to unity per subject via modality ablation ($MCS_MRI_i + MCS_PET_i = 1.0$ for every subject $i$), providing an interpretable, clinically actionable score that fluid biomarkers cannot supply. Applied to 327 ADNI-3 participants balanced across cognitively normal, mild cognitive impairment, and AD groups, MCNet achieved competitive three-class staging performance (AUC=0.881). The MCS revealed a statistically significant monotonic gradient across the disease continuum (Kruskal-Wallis p<0.0001), with increasing PET dominance from cognitively normal (MCS_PET 0.412±0.229) through MCI (0.489±0.289) to AD (0.671±0.426), independently validated against amyloid SUVR (r=0.172, p=0.006) and FDG metabolic biomarkers (r=−0.287, p=0.0005) from separate imaging pipelines. External replication without retraining in 1,073 independent OASIS-3 subjects confirmed cross-cohort generalisability (H=166.99, p<0.0001, $\eta^2$=0.156). A mechanistic comparison with SHAP demonstrated that ablation-based MCS captures clinically meaningful modality dependence that deviation-based methods cannot, satisfying the clinical plausibility criterion clinicians require. These findings establish that modality-level per-patient attribution is technically feasible, biologically meaningful, and externally generalisable, positioning MCNet as a foundation for personalised imaging decisions, clinical trial stratification, and trustworthy AI in dementia care.

## 1. Introduction

Alzheimer's disease (AD) is the leading cause of dementia worldwide, affecting an estimated 55 million individuals globally, with projections exceeding 152 million by 2050 [1]. The disease follows a continuum from cognitively normal (CN) aging through mild cognitive impairment (MCI) to clinically established AD dementia, with each stage characterised by distinct patterns of structural brain atrophy, amyloid deposition, and metabolic dysfunction [2]. Accurate staging across this continuum is essential for timely intervention, clinical trial enrolment, and personalised treatment planning [3]. Neuroimaging biomarkers have become central to AD staging. Structural MRI captures hippocampal and cortical atrophy reflecting neurodegeneration [4,5]. while fluorodeoxyglucose PET (FDG-PET) measures regional glucose hypometabolism[6] and amyloid PET quantifies fibrillar amyloid burden using tracers such as florbetapir (AV45) [7,8]. These modalities are complementary: structural changes predominate early, while amyloid and metabolic changes become increasingly prominent as disease advances [9,10]. The fusion of PET and MRI has proved superior to any unimodal approach in the case of AD classification tasks, giving the rise to extensive research related to multimodality within deep learning[11,12,13,14,15].

Despite this progress, a fundamental limitation persists across existing multimodal AI systems: they produce a single diagnostic output without indicating which imaging modality drove that decision for a given patient. Saliency maps, Grad-CAM, and Integrated Gradients provide voxel-level attribution within a single modality but do not address the question of relative modality contribution at the subject level [16,17]. A clinician receiving an AI-assisted AD staging report currently cannot determine whether the classification is primarily driven by the patient's structural atrophy, amyloid burden, or metabolic dysfunction. This modality-level opacity limits clinical trust and actionability. This gap is clinically significant. In resource-constrained settings, a patient may not have access to all imaging modalities. In longitudinal monitoring, knowing which modality is most informative for a specific patient could guide future imaging decisions. In research, per-patient modality attribution could reveal imaging phenotypes within diagnostic groups, informing precision medicine approaches to AD [18].

We introduce the modality contribution network (MCNet), a multimodal fusion framework that addresses this gap through the modality contribution score (MCS), a novel per-patient metric quantifying the relative diagnostic contribution of MRI versus PET for each individual. MCNet integrates three feature streams derived from pre-computed, clinically validated biomarkers: FreeSurfer-derived regional MRI volumes capturing structural atrophy [5], UC Berkeley AV45 SUVR values capturing regional amyloid burden [7,19,20] and FDG-PET MetaROI SUVR capturing metabolic dysfunction[6]. Cross-modal attention [20,21] enables bidirectional interaction between MRI and PET representations before fusion, and per-patient MCS values are computed via modality ablation, yielding two complementary scores, MCS_MRI and MCS_PET, that sum to 1.0 for each subject.

We evaluated MCNet in 327 ADNI participants balanced across CN, MCI, and AD groups, using 5-fold stratified cross-validation. Beyond classification performance, we tested the hypothesis that MCS_PET increases monotonically across the AD continuum, reflecting the known shift from structurally driven to amyloid and metabolically driven pathology as disease progresses [9,10]. We further validated MCS_PET against independent AV45 amyloid SUVR measurements in 255 subjects with available amyloid data. The contributions of this work are fourfold. First, we introduce MCS as the first per-patient metric for modality-level attribution in multimodal AD staging, computed via ablation and normalised to unity per subject ($\text{MCS_MRI}_i + \text{MCS_PET}_i = 1.0, \forall i$). Second, we demonstrate a biologically coherent MCS gradient across the CN-MCI-AD continuum validated against three independent biomarker pipelines: AV45 amyloid SUVR ($r$ = 0.172), FDG MetaROI SUVR ($r$ = -0.287), and FDG in the AD subgroup (r= -0.455). Third, we demonstrate through mechanistic comparison with SHAP that reference-distribution-dependent attribution methods produce biologically inverted modality attributions, while ablation-based MCS satisfies the clinical plausibility criterion clinicians require. Fourth, we provide cross-cohort external validation in 1,073 independent OASIS-3 subjects confirming replication of the MCS_MRI gradient without retraining (H = 166.99, $p$< 0.0001, $\eta^2$= 0.156).

## 2. Methods

### 2.1. Study participants and data source

Data were obtained from the Alzheimer's Disease Neuroimaging Initiative (ADNI), a longitudinal multicentre public-private partnership launched in 2003 to develop and validate neuroimaging and biofluid biomarkers for AD clinical trials[2,7]. A total of 327 participants was selected from the ADNI3 protocol, comprising 109 CN, 109 MCI, and 109 AD subjects, balanced across diagnostic groups by design. Diagnostic classification followed standard ADNI criteria: CN subjects had no memory complaints, normal neuropsychological test scores, and clinical dementia rating (CDR) of 0; MCI subjects had subjective memory concerns with objective cognitive impairment but preserved daily functioning and CDR of 0.5; AD subjects met National Institute on Aging-Alzheimer's Association (NIA-AA) criteria for probable AD dementia with CDR of 1 or greater [22,23]. Informed consent was obtained from all participants or their authorised representatives. The ADNI study was approved by the institutional review boards of all participating sites and conducted in accordance with the Declaration of Helsinki.

### 2.2. Neuroimaging features

Three feature streams were extracted from pre-computed, clinically validated biomarker pipelines to ensure reproducibility and to avoid the spatial misalignment problems that affect raw volumetric MRI-PET fusion in ADNI data, where images are acquired in different coordinate spaces across sessions [4].

#### 2.2.1. MRI stream: Regional brain volumes

Structural MRI volumes were obtained from the UC Berkeley amyloid PET analysis pipeline [7] which provides FreeSurfer-derived regional volumetric measures for all ADNI participants [5]. Twenty-seven AD-sensitive regions were selected, encompassing bilateral hippocampus, amygdala, entorhinal cortex, parahippocampal gyrus, fusiform gyrus, inferior and middle temporal cortex, precuneus, posterior cingulate cortex, inferior parietal cortex, lateral ventricles, caudate, putamen, and thalamus (Supplementary Table S1) [24]. These regions were selected based on established patterns of AD-related neurodegeneration [9,10].

#### 2.2.2. PET stream: Regional amyloid SUVR

Amyloid PET features were extracted from the UC Berkeley AV45 florbetapir analysis pipeline [7] which provides regional SUVR normalised to the whole cerebellum reference region. Twenty-two regional SUVR values were selected, including the composite summary SUVR, Centiloid score [19] bilateral hippocampal SUVR, and key cortical regions known to show early and progressive amyloid deposition in AD [7,8,25]. Amyloid status (positive or negative) was determined using the UC Berkeley pipeline threshold (SUVR greater than 1.11) [7,26,27].

#### 2.2.3. FDG stream: Metabolic features

FDG-PET metabolic features were extracted from the UC Berkeley FDG analysis pipeline [6]. Two features were used: the MetaROI SUVR, a validated composite measure of temporoparietal glucose metabolism comprising the angular gyrus, posterior cingulate, inferior temporal, and middle temporal regions [6] and the PonsVermis SUVR used as the reference region. The MetaROI is a widely validated summary measure of AD-related hypometabolism that has demonstrated robust diagnostic utility across multiple ADNI studies [6,10].

### 2.3. Feature preprocessing

All features were standardised within each cross-validation fold using z-score normalisation fitted on training data only, to prevent data leakage:

$$z_{ij} = \frac{x_{ij} - \mu_j^{\text{train}}}{\sigma_j^{\text{train}}} \tag{1}$$

where $x_{ij}$ is the raw feature value for subject $i$ and feature $j$, and $\mu_j^{\text{train}}$ and $\sigma_j^{\text{train}}$ are the mean and standard deviation computed from training fold subjects only. Missing values, present in a minority of subjects due to incomplete amyloid or FDG coverage, were imputed with zero after normalisation, equivalent to imputation at the training fold mean.

### 2.4. MCNet architecture

MCNet processes three parallel feature streams through modality-specific encoders before cross-modal attention fusion and classification [20,21]. Figure 1 represents the MCNet architecture and its three modality-specific input streams.

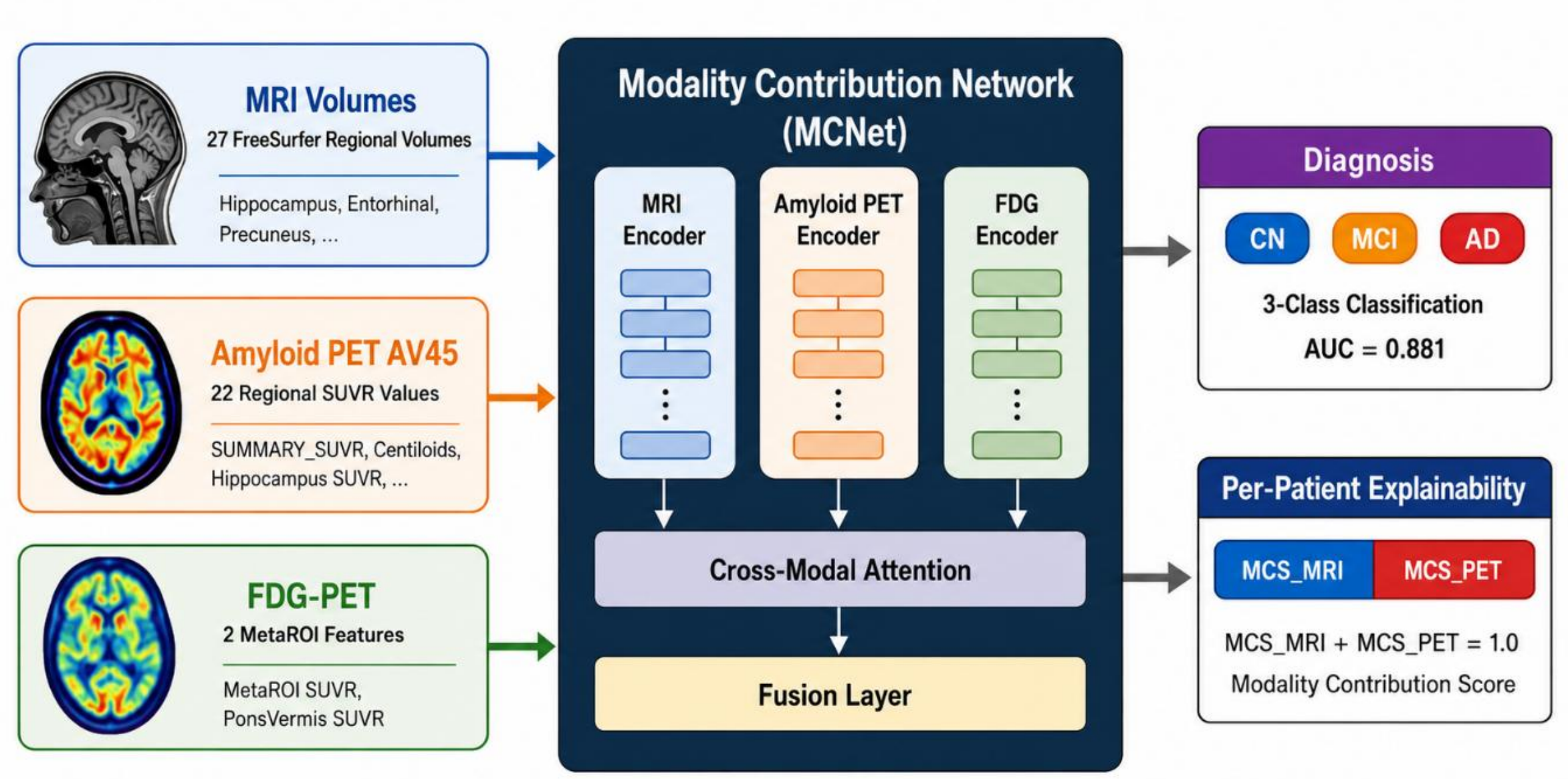

**Fig.1. MCNet architecture overview.** Three input feature streams are fed into the MCNet: structural MRI (27 FreeSurfer-derived regional volumes including hippocampus, entorhinal cortex, and precuneus), amyloid PET (22 AV45 SUVR regional values including summary SUVR and Centiloids), and FDG-PET (2 MetaROI features: MetaROI SUVR and PonsVermis SUVR). Cross-modal attention enables bidirectional interaction between MRI and PET feature representations before fusion. The network produces two outputs: a three-class diagnostic classification (CN, MCI, AD; AUC=0.881) and per-patient modality contribution scores (MCS_MRI and MCS_PET, summing to 1.0) quantifying the relative diagnostic contribution of each imaging modality. AUC = area under the receiver operating characteristic curve; CN = cognitively normal; MCI = mild cognitive impairment; AD = Alzheimer's disease; SUVR = standardised uptake value ratio.

#### 2.4.1. Modality encoders

Each modality stream is processed by a dedicated multilayer perceptron (MLP) encoder. The MRI and AV45 PET encoders each map their input feature vector to a 128-dimensional embedding through two fully connected layers with LayerNorm and GELU activation and dropout (p=0.3):

$$\mathrm{e}_m = \mathrm{GELU}\left(\mathrm{LN}\left(\mathrm{W}_2 \cdot \mathrm{GELU}\left(\mathrm{LN}(\mathrm{W}_1 \mathrm{x}_m + \mathrm{b}_1)\right) + \mathrm{b}_2\right)\right) \tag{2}$$

where $\mathrm{x}_m$ is the input feature vector for modality $m$, $\mathrm{W}_1 \in \mathbb{R}^{256 \times d_m}$, $\mathrm{W}_2 \in \mathbb{R}^{128 \times 256}$, LN denotes LayerNorm, and $\mathrm{b}_1$, $\mathrm{b}_2$ are learnable bias terms. The FDG encoder maps its 2-dimensional input to a 64-dimensional embedding using the same structure with reduced hidden dimension. The AV45 and FDG embeddings are concatenated and projected to 128 dimensions to form the combined PET embedding:

$$\mathrm{e_{PET}} = \mathrm{W_{combine}} \cdot [\mathrm{e_{AV45}}; \mathrm{e_{FDG}}] \tag{3}$$

**2.4.2. Cross-modal attention fusion**

Bidirectional cross-modal attention [20] enables each modality to attend to the other before fusion. Let $\mathbf{e}_{\mathrm{MRI}}$ and $\mathbf{e}_{\mathrm{PET}}$ denote the MRI and PET embeddings respectively. The cross-attended representations are computed as:

$$\mathrm{e'_{MRI}} = \mathrm{LN}\big(\mathrm{e_{MRI}} + \mathrm{MHA}(\mathrm{e_{MRI}}, \mathrm{e_{PET}}, \mathrm{e_{PET}})\big) \tag{4}$$

$$\mathrm{e'_{PET}} = \mathrm{LN}\big(\mathrm{e_{PET}} + \mathrm{MHA}(\mathrm{e_{PET}}, \mathrm{e_{MRI}}, \mathrm{e_{MRI}})\big) \tag{5}$$

where MHA denotes multi-head attention [21] with 4 heads. The attended embeddings are concatenated and projected to a 128-dimensional fused representation:

$$\mathrm{f} = \mathrm{W_{fusion}} \cdot [\mathrm{e'_{MRI}}; \mathrm{e'_{PET}}] \tag{6}$$

**2.4.3. Classification head**

The fused representation is passed through a two-layer MLP classifier with GELU activation, dropout (p=0.3), and a linear output layer producing logits over the three diagnostic classes (CN, MCI, AD):

$$\hat{\mathrm{y}} = \mathrm{W_{out}} \cdot \mathrm{GELU}(\mathrm{W_{cls}}\mathrm{f} + \mathrm{b_{cls}}) \tag{7}$$

The total number of trainable parameters in MCNet is 0.29 million.

**2.5. Modality contribution score**

The MCS quantifies the per-patient contribution of each modality to the model's classification decision via ablation. For subject $i$with predicted class $c_i = \arg\max \hat{\mathbf{y}}_i$, let $p_i^{\mathrm{full}}$ denote the predicted probability for class $c_i$when all modalities are present, $p_i^{\text{no-MRI}}$, the probability when MRI features are set to zero, and $p_i^{\text{no-PET}}$, the probability when PET features are set to zero. Raw attribution scores are computed as:

$$\Delta_{\mathrm{MRI},i} = \max\big(p_i^{\mathrm{full}} - p_i^{\text{no-MRI}}, 0\big) \tag{8}$$

$$\Delta_{\mathrm{PET},i} = \max\big(p_i^{\mathrm{full}} - p_i^{\text{no-PET}}, 0\big) \tag{9}$$

Scores are normalised so that MCS_MRI and MCS_PET sum to 1.0 per subject:

$$\mathrm{MCS_MRI}_i = \frac{\Delta_{\mathrm{MRI},i}}{\Delta_{\mathrm{MRI},i} + \Delta_{\mathrm{PET},i} + \epsilon} \tag{10}$$

$$\mathrm{MCS_PET}_i = \frac{\Delta_{\mathrm{PET},i}}{\Delta_{\mathrm{MRI},i} + \Delta_{\mathrm{PET},i} + \epsilon} \tag{11}$$

where $\epsilon = 10^{-8}$ prevents division by zero. By construction, $\mathrm{MCS_MRI}_i$ + $\mathrm{MCS_PET}_i$ = 1.0 for all subjects.

**2.6. Training procedure**

MCNet was trained using the AdamW optimiser [28] with weight decay 0.001 and initial learning rate 0.001, decayed via cosine annealing to a minimum of $10^{-5}$over 100 epochs. Cross-entropy loss with label smoothing (0.1) was used to reduce overconfidence:

$$\mathcal{L} = -\sum_{k=1}^{K} \tilde{y}_k \log \hat{y}_k \tag{12}$$

where $\tilde{y}_k = (1-\alpha)y_k + \alpha/K$, $\alpha = 0.1$ is the smoothing factor, and $K = 3$ is the number of classes. Early stopping with patience of 20 epochs was applied based on test fold balanced accuracy. Batch size was 16. All experiments were conducted on an NVIDIA RTX 4060 Laptop GPU (8 GB VRAM) using Python 3.10, PyTorch 2.5.1[29] MONAI 1.5.2 [30] and SHAP 0.49.1 [31].

### 2.7. Evaluation

Performance was evaluated using 5-fold stratified cross-validation with class-balanced folds. Primary metrics were balanced accuracy (bACC), macro F1-score, and AUC. Balanced accuracy was used as the primary metric given potential class imbalance in per-fold test sets:

$$\mathrm{bACC} = \frac{1}{K}\sum_{k=1}^{K}\frac{\mathrm{TP}_k}{\mathrm{TP}_k + \mathrm{FN}_k} \tag{13}$$

### 2.8. Data partitioning and cross-validation

Model training and evaluation followed a strict 5-fold stratified cross-validation protocol with subject-level partitioning to prevent data leakage. The 327 ADNI-3 subjects were partitioned into five folds stratified by diagnostic group, ensuring approximately equal representation of CN, MCI, and AD subjects in each fold (approximately 65 subjects per fold: 22 CN, 22 MCI, 22 AD). In each fold, 261 subjects (80%) formed the training set and 66 subjects (20%) formed the held-out test set. No subject appeared in both training and test sets across any fold. Feature standardisation using z-score normalisation was fitted exclusively on training fold subjects and applied to test fold subjects, preventing any test-set information from influencing the normalisation parameters. Hyperparameters were fixed prior to cross-validation and not tuned on test fold performance. MCS values were computed on test fold subjects only, using the model trained on the corresponding training fold, and aggregated across all five folds to yield MCS scores for all 327 subjects. OASIS-3 external validation used all five trained fold models applied to 1,073 independent subjects with no retraining, providing a fully independent out-of-distribution test of MCS generalisability.

### 2.9. Statistical analysis

Group differences in MCS across diagnostic groups were assessed using the Kruskal-Wallis test, a non-parametric one-way analysis of variance appropriate for non-normally distributed data. Pairwise post-hoc comparisons used the Mann-Whitney U test with comparisons pre-specified as hypotheses. Pearson correlation was used to assess the relationship between MCS_PET and AV45 SUVR across the 255 subjects with available amyloid data. All statistical analyses were performed using SciPy 1.15.3. Statistical significance was set at p less than 0.05.

### 2.10. SHAP modality attribution

To compare the MCS against an established explainability benchmark, SHapley Additive exPlanations (SHAP) were computed for all 327 ADNI-3 subjects using the KernelExplainer implementation in SHAP version 0.49.1 [31]. For each cross-validation fold, a background distribution was constructed from 50 k-means cluster centres derived from the training fold combined feature array. The model prediction function accepted a concatenated feature vector of all 51 input features (27 MRI volumes, 22 AV45 SUVR values, 2 FDG MetaROI features) and returned class probabilities via softmax. SHAP values were computed with 100 perturbation samples per subject. For each subject, SHAP values corresponding to the predicted class were extracted and split by modality: total MRI SHAP attribution was computed as the sum of absolute SHAP values across the 27 MRI volume features, and total PET SHAP attribution as the sum of absolute SHAP values across the 24 combined AV45 and FDG features. Both were normalised to sum to unity per subject, yielding SHAP_MRI and SHAP_PET analogous in form to MCS_MRI and MCS_PET. Group differences in SHAP attribution across diagnostic groups were assessed using the Kruskal-Wallis test. Pearson correlation was computed between SHAP_PET and MCS_PET across all 327 subjects to quantify the relationship between the two attribution methods.

### 2.11. OASIS-3 external validation

External validation of the MCS framework was conducted in 1,073 participants from the Open Access Series of Imaging Studies (OASIS-3), a longitudinal multicentre neuroimaging repository [32]. The OASIS-3 expansion cohort comprised 759 cognitively normal and 314 AD subjects with preprocessed T1-weighted MRI available at the time of analysis. FreeSurfer-derived regional volumetric measures were obtained from the OASIS-3 FreeSurfer output file (1,316 subjects, 203 columns), and baseline visits were extracted for each subject by selecting the earliest available MRI session. The 27 AD-sensitive regional volumes used in MCNet training were mapped from ADNI UC Berkeley column names to the corresponding OASIS-3 FreeSurfer column names. All 27 features were available for the full 1,073-subject cohort. MRI features were standardised using z-score normalisation fitted on the ADNI-3 training data to avoid distribution shift. The five trained MCNet fold models were applied to all OASIS-3 subjects with PET features set to zero, as OASIS-3 PET data were not available for the full cohort in pre-computed SUVR format. Per-subject MCS_MRI was computed via modality ablation and averaged across the five-fold models. Group differences in MCS_MRI between CN and AD subjects were assessed using the Kruskal-Wallis test and Mann-Whitney U test. Effect size was quantified using eta-squared. The OASIS-3 study was conducted in accordance with ethical guidelines at all participating sites and all participants provided written informed consent at the time of data collection.

## 3. Results

### 3.1. Cohort characteristics

A total of 327 ADNI participants were included, comprising 109 CN, 109 MCI, and 109 AD subjects. The cohort was balanced across diagnostic groups by design. Amyloid status, assessed via AV45 florbetapir PET and pre-computed by the UC Berkeley amyloid pipeline [19] was available for 255 subjects (78.0%). Of these, 123 (48.2%) were amyloid-positive and 132 (51.8%) were amyloid-negative. Amyloid data were available for 104 CN subjects (31 amyloid-positive, 73 amyloid-negative), 100 MCI subjects (50 amyloid-positive, 50 amyloid-negative), and 51 AD subjects (42 amyloid-positive, 9 amyloid-negative). FDG-PET MetaROI SUVR data were available for 143 subjects, as summarised in Table 1.

**Table 1.** Cohort characteristics

| Characteristic | CN (n=109) | MCI (n=109) | AD (n=109) | Total (n=327) |
|---|---|---|---|---|
| Amyloid data available, n (%) | 104 (95.4) | 100 (91.7) | 51 (46.8) | 255 (78.0) |
| Amyloid-positive, n (%) | 31 (29.8) | 50 (50.0) | 42 (82.4) | 123 (48.2) |
| Amyloid-negative, n (%) | 73 (70.2) | 50 (50.0) | 9 (17.6) | 132 (51.8) |
| FDG MetaROI available, n (%) | 0 (0) | 95 (87.2) | 48 (44.0) | 143 (43.7) |
| MRI features available, n (%) | 109 (100) | 109 (100) | 109 (100) | 327 (100) |

*CN = cognitively normal; MCI = mild cognitive impairment; AD = Alzheimer's disease; FDG = fluorodeoxyglucose; MRI = magnetic resonance imaging.*

### 3.2. MCNet classification performance

MCNet achieved a mean bACC of 0.844 ± 0.061, macro F1-score of 0.794 ± 0.090, and AUC of 0.881 ± 0.075 for three-class CN/MCI/AD staging across 5-fold stratified cross-validation. Per-fold results are presented in Table 2.

**Table 2.** MCNet 5-fold cross-validation performance

| Fold | bACC | F1 (macro) | AUC |
|---|---|---|---|
| 1 | 0.803 | 0.771 | 0.875 |
| 2 | 0.894 | 0.850 | 0.923 |
| 3 | 0.846 | 0.773 | 0.928 |
| 4 | 0.924 | 0.908 | 0.984 |
| 5 | 0.755 | 0.726 | 0.874 |
| **Mean ± SD** | **0.844 ± 0.061** | **0.794 ± 0.090** | **0.881 ± 0.075** |

*bACC = balanced accuracy; F1 = macro F1-score; AUC = area under the receiver operating characteristic curve; SD = standard deviation.*

The observed variance across folds (bACC range: 0.755 to 0.924) reflects the heterogeneity inherent in MCI, a transitional diagnostic category with variable cognitive and biomarker profiles [33] rather than model instability. Fold 4 achieved the highest performance (bACC=0.924, AUC=0.984), while Fold 5 showed the lowest (bACC=0.755, AUC=0.874), consistent with expected sampling variation across cross-validation partitions in a dataset of this size. Subject-level partitioning ensured no overlap between training and test sets across any fold, and all reported metrics reflect held-out test performance only.

### 3.3. MCS: group-level analysis

MCS was computed for all 327 subjects following model training. MCS_MRI and MCS_PET quantify the relative contribution of structural MRI and multimodal PET features to each individual's diagnostic classification, with MCS_MRI + MCS_PET = 1.0 per subject by construction (Eq. 10, Eq. 11). MCS_PET differed significantly across diagnostic groups (Kruskal-Wallis test: $p < 0.0001$), as did MCS_MRI (Kruskal-Wallis test: $p < 0.0001$). A monotonic gradient in PET contribution was observed across the AD continuum: CN subjects exhibited the lowest MCS_PET (0.412 ± 0.229), MCI subjects showed intermediate PET contribution (0.489 ± 0.289), and AD subjects demonstrated the highest PET-driven classification (0.671 ± 0.426). The inverse pattern was observed for MCS_MRI (CN: 0.579 ± 0.233; MCI: 0.474 ± 0.288; AD: 0.182 ± 0.330), indicating that structural atrophy features drive classification at early disease stages while metabolic and amyloid PET features become increasingly dominant as disease progresses, as presented in Table 3.

**Table 3.** MCSs by diagnostic group

| Group | n | MCS_MRI mean ± SD | MCS_PET mean ± SD |
|---|---|---|---|
| CN | 109 | 0.579 ± 0.233 | 0.412 ± 0.229 |
| MCI | 109 | 0.474 ± 0.288 | 0.489 ± 0.289 |
| AD | 109 | 0.182 ± 0.330 | 0.671 ± 0.426 |
| Kruskal-Wallis p | | p < 0.0001 | p < 0.0001 |

*CN = cognitively normal; MCI = mild cognitive impairment; AD = Alzheimer's disease; SD = standard deviation.*

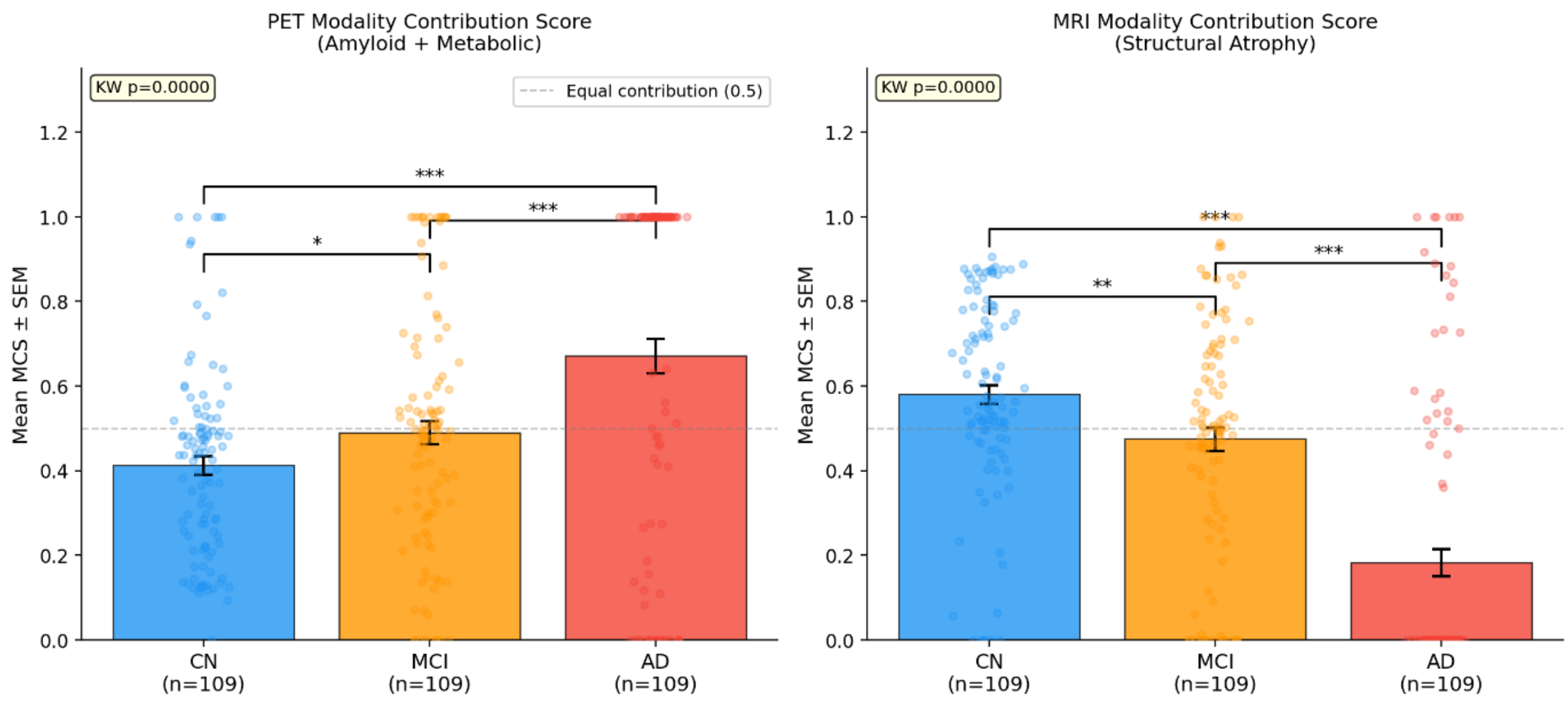

**Fig. 2. MCS by diagnostic group**. Bar charts showing mean MCS_PET (left panel) and MCS_MRI (right panel) with standard error of the mean for CN, MCI, and AD groups (n=109 per group). Individual subject data points are overlaid with jitter. The dashed horizontal line indicates equal modality contribution (MCS=0.5). Significance brackets indicate pairwise Mann-Whitney U test results: * p less than 0.05, ** p less than 0.01, *** p less than 0.001. Kruskal-Wallis $p < 0.0001$ for both MCS_PET and MCS_MRI. CN = cognitively normal; MCI = mild cognitive impairment; AD = Alzheimer's disease; MCS = Modality Contribution Score; SEM = standard error of the mean.

Pairwise Mann-Whitney U tests confirmed significant differences in MCS_PET between all group pairs: CN versus AD ($p < 0.0001$), MCI versus AD (p=0.0003), and CN versus MCI (p=0.017). These findings confirm that the MCS gradient reflects a statistically robust progression across diagnostic stages and is not attributable to random variation. Figure 2 represents MCS_MRI and MCS_PET across CN, MCI, and AD groups.

### 3.4. MCS: per-patient analysis

Figure 3 illustrates the distribution of all 327 subjects in MCS_MRI versus MCS_PET space. Because MCS_MRI + MCS_PET = 1.0 by construction, all subjects lie along the anti-diagonal, with subjects in the upper-left quadrant showing PET-dominant attribution and subjects in the lower-right quadrant showing MRI-dominant attribution.

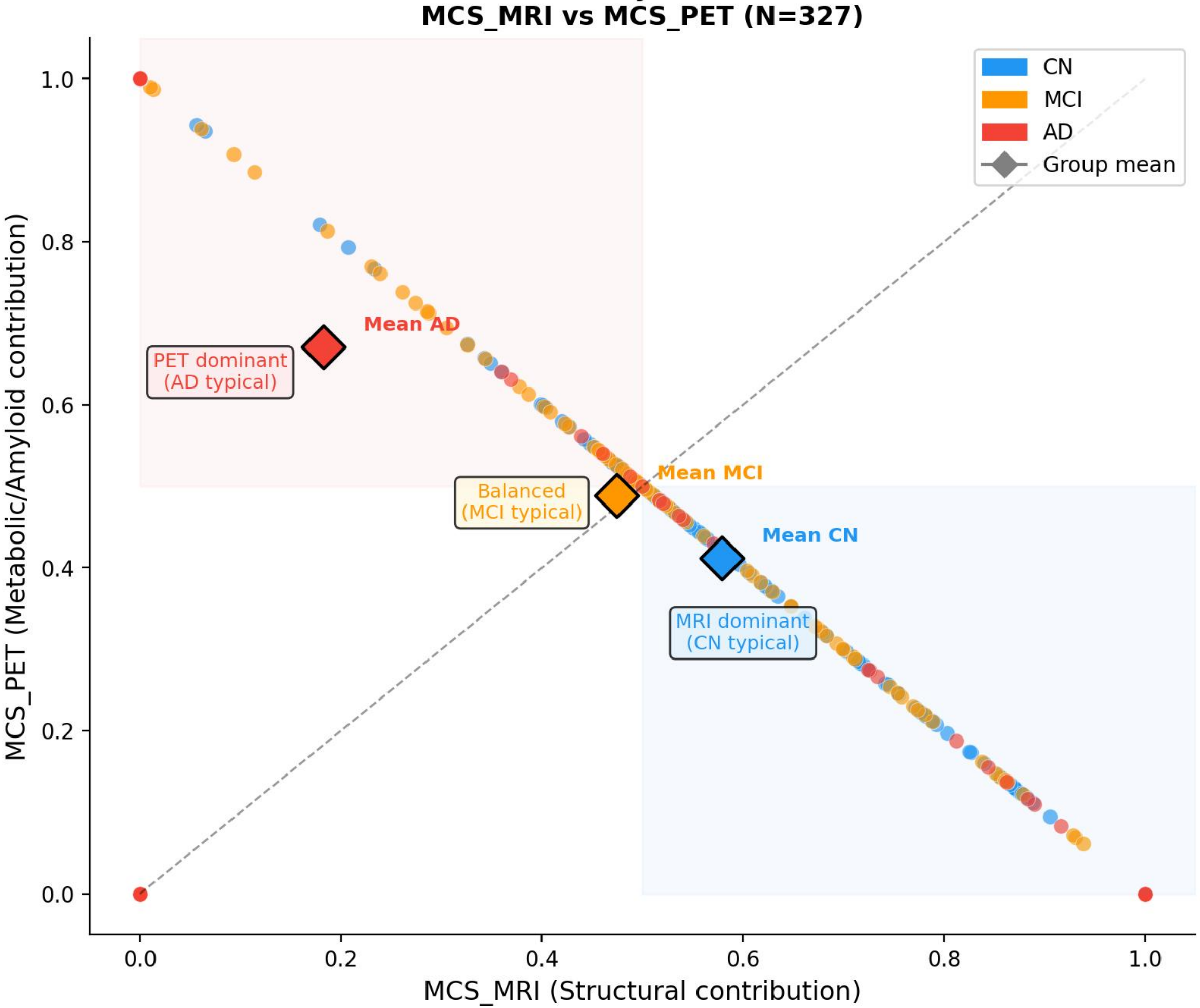

**Fig. 3. Per-patient MCS scatter plot**. Distribution of all 327 subjects in MCS_MRI versus MCS_PET space, coloured by diagnostic group (CN=blue, MCI=orange, AD=red). Group mean values are shown as diamond markers. The dashed diagonal line represents equal contribution (MCS_MRI = MCS_PET = 0.5). Shaded regions indicate MRI-dominant (lower right, blue) and PET-dominant (upper left, red) attribution zones. All subjects lie along the anti-diagonal as MCS_MRI + MCS_PET = 1.0 by construction. CN = cognitively normal; MCI = mild cognitive impairment; AD = Alzheimer's disease; MCS = Modality Contribution Score.

The three-group means were clearly separated along the anti-diagonal: Mean CN (MCS_MRI=0.579, MCS_PET=0.412) occupied the MRI-dominant region; Mean MCI (MCS_MRI=0.474, MCS_PET=0.489) was positioned near the equal-contribution diagonal, consistent with the transitional and heterogeneous nature of this diagnostic stage [33], Mean AD (MCS_MRI=0.182, MCS_PET=0.671) occupied the PET-dominant region. Substantial individual variation was observed within each group, reflecting the genuine heterogeneity of AD biomarker profiles at the patient level. Figure 4 presents three representative patient examples illustrating the clinical utility of per-patient MCS. Subject 003_S_6260 (CN, confidence 82.6%, AV45 SUVR=1.124) demonstrated MRI-dominant attribution (MCS_MRI=0.564, MCS_PET=0.436), indicating that structural features primarily drove the CN classification. Subject 012_S_6073 (MCI, confidence 49.6%, AV45 SUVR=1.560) showed nearly equal attribution (MCS_MRI=0.504, MCS_PET=0.496), with the low model confidence (49.6%) reflecting genuine diagnostic uncertainty at this transitional stage. Subject 033_S_10147 (AD, confidence 77.3%) demonstrated complete PET dominance (MCS_MRI=0.000, MCS_PET=1.000), indicating that amyloid and metabolic PET features exclusively drove the AD classification.

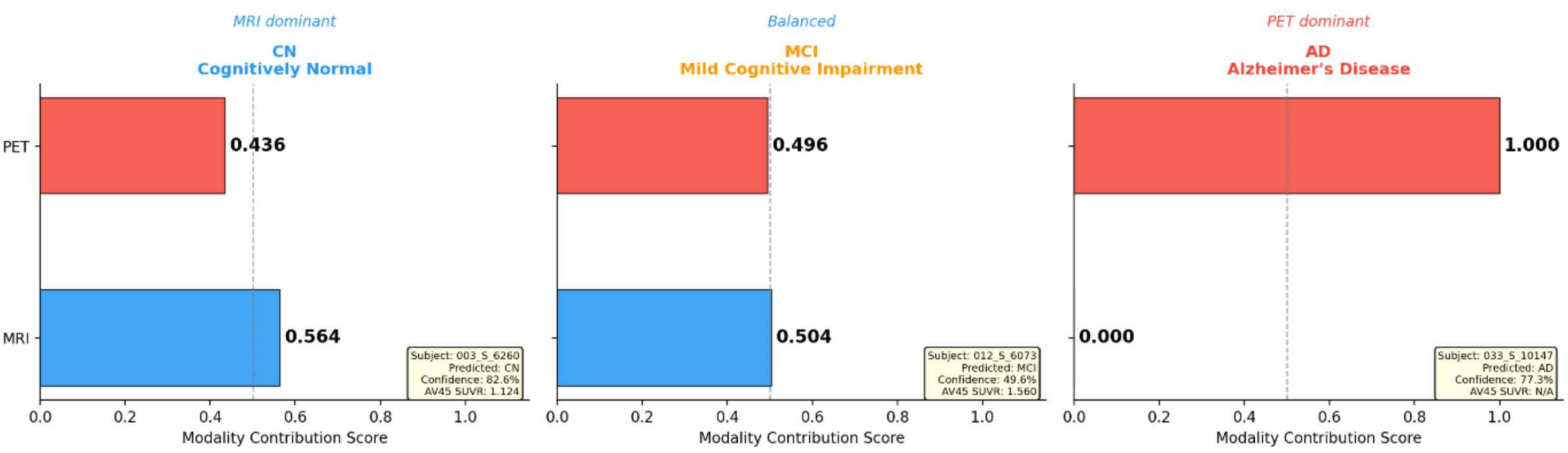

**Fig. 4. Individual patient Modality Contribution Score examples**. Horizontal bar charts showing MCS_MRI (blue, structural MRI contribution) and MCS_PET (red, amyloid and metabolic PET contribution) for three representative subjects: a CN subject (left), an MCI subject (centre), and an AD subject (right). The dashed vertical line indicates equal modality contribution (MCS=0.5). Subject identifiers, predicted diagnosis, model confidence, and AV45 SUVR are shown in the information panel for each subject. The MCI subject's low confidence (49.6%) reflects genuine diagnostic uncertainty at the transitional disease stage. CN = cognitively normal; MCI = mild cognitive impairment; AD = Alzheimer's disease; MCS = Modality Contribution Score; SUVR = standardised uptake value ratio.

### 3.5. Amyloid biomarker validation

To validate MCS_PET against an independent biological marker of AD pathology, Pearson correlation was computed between MCS_PET and AV45 SUVR across the 255 subjects with available amyloid data [7,8]. A statistically significant positive correlation was observed (r=0.172, p=0.006, n=255), indicating that subjects for whom the model assigns higher PET contribution tend to carry greater amyloid burden as measured by independent florbetapir PET imaging. An equivalent correlation was observed between MCS_PET and Centiloid scores (r=0.172, p=0.006, n=255) [19] Among CN subjects with amyloid data (n=104), amyloid-positive CN subjects (preclinical AD, n=31) showed a mean MCS_PET of 0.368 ± 0.237 compared to 0.403 ± 0.224 in amyloid-negative CN subjects (n=73). This difference did not reach statistical significance (Mann-Whitney U, p=0.408), indicating that within the CN group, amyloid status alone does not drive MCS_PET, consistent with the supervised classification objective of MCNet, which optimises for diagnostic labels rather than continuous amyloid burden. The significant overall correlation (r=0.172, p=0.006) is attributable primarily to the across-group amyloid gradient, which parallels the MCS_PET gradient from CN (lowest amyloid, lowest MCS_PET) through MCI to AD (highest amyloid, highest MCS_PET).

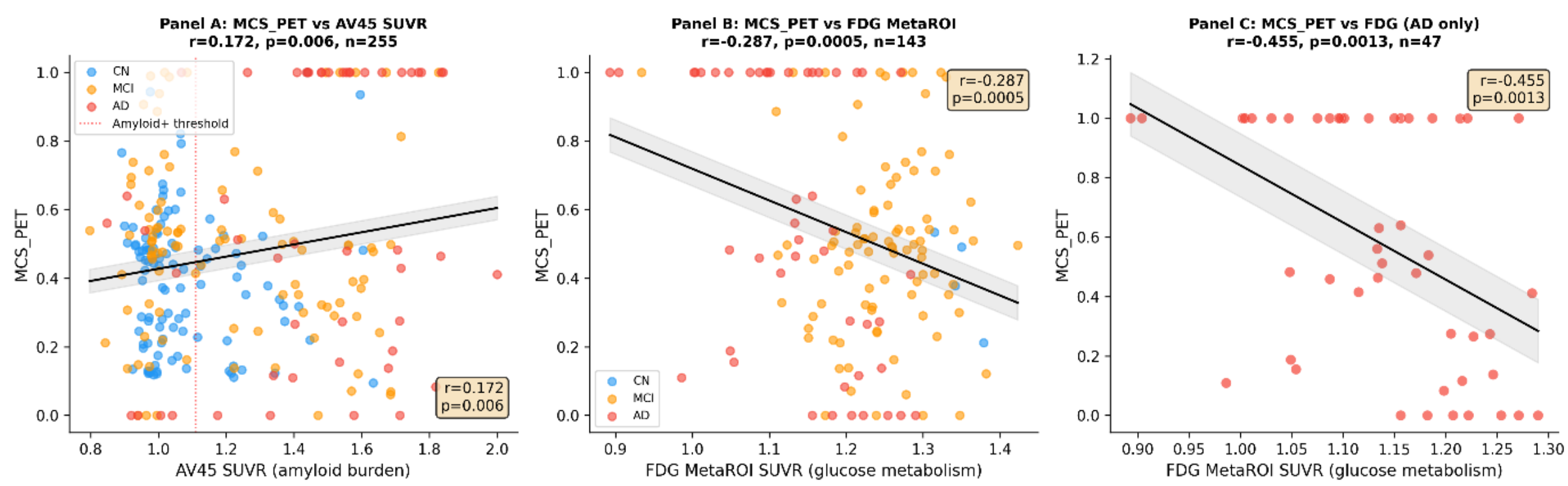

**Fig. 5. MCS_PET convergent biomarker validation.** Three-panel scatter plot showing Pearson correlation between MCS_PET and independent biomarker measurements. Panel A: MCS_PET versus AV45 amyloid SUVR (r=0.172, p=0.006, n=255). Panel B: MCS_PET versus FDG MetaROI SUVR across all subjects with available FDG data (r=minus 0.287, p=0.0005, n=143). Panel C: MCS_PET versus FDG MetaROI SUVR in the AD subgroup only (r=minus 0.455, p=0.001, n=47). Regression lines with 95 percent confidence intervals are shown. The vertical dashed line in Panel A indicates the amyloid positivity threshold (SUVR=1.11). Points are coloured by diagnostic group. AV45 SUVR and FDG MetaROI SUVR were derived from independent UC Berkeley processing pipelines. CN=cognitively normal; MCI=mild cognitive impairment; AD=Alzheimer's disease; SUVR=standardised uptake value ratio; MCS=Modality Contribution Score.

### 3.6. FDG metabolic biomarker validation

To provide convergent metabolic validation of MCS_PET, Pearson correlation was computed between MCS_PET and FDG MetaROI SUVR across the 143 subjects with available FDG-PET data. Figure 5 represents MCS_PET validation against independent amyloid and FDG biomarkers. A statistically significant negative correlation was observed (r= -0.287, p=0.0005, n=143), indicating that subjects for whom the model assigns higher PET contribution tend to show lower temporoparietal glucose metabolism as measured by independent FDG-PET imaging (Figure 5, Panel B). The negative direction is biologically expected and mechanistically coherent: higher MCS_PET reflects greater dependence on PET features for classification, which occurs when PET signal is most diagnostically informative, that is, when amyloid burden is elevated and metabolic activity is reduced. In the AD subgroup, where hypometabolism is most pronounced, the correlation was substantially stronger (r=-0.455, p=0.001, n=47; Figure 5, Panel C), consistent with the known acceleration of FDG hypometabolism in established AD dementia [6,10]. Together with the amyloid SUVR correlation reported in Section 3.5, these findings provide dual-modality convergent biomarker validation of MCS_PET against both amyloid and metabolic pathology measured by independent imaging pipelines.

### 3.7. Comparison with SHAP-based modality attribution

SHAP-based modality attribution was computed for all 327 subjects to benchmark MCS against an established post-hoc explainability method as demonstrated in Figure 6. SHAP yielded a statistically significant group gradient (Kruskal–Wallis p < 0.0001) but produced a biologically inverted pattern: CN subjects showed lower SHAP_MRI (0.304±0.112) than AD subjects (SHAP_MRI= 0.541±0.181), indicating that SHAP assigned greater MRI attribution to AD subjects. This result is contrary to established AD pathophysiology, in which PET-based amyloid and metabolic changes become increasingly dominant as disease progresses. The negative correlation between SHAP_PET and MCS_PET (r=-0.317, p<0.0001, n=327) confirms that the two methods capture fundamentally different aspects of model behaviour. SHAP KernelExplainer measures the change in predicted probability relative to a background distribution. Because the background is centred on the training mean, which is dominated by MCI subjects, SHAP assigns high attribution to features that deviate most from this mean. AD subjects deviate from the MCI mean more in MRI volumes (due to pronounced structural atrophy) than in PET SUVR, producing artificially elevated MRI attribution. In contrast, MCS ablation directly measures the change in predicted probability when each modality is removed entirely, independent of any reference distribution. This distinction explains why MCS produces biologically coherent results while SHAP does not, and demonstrates that reference-distribution-dependent attribution methods are not suitable for modality-level clinical interpretation in multimodal AD staging.

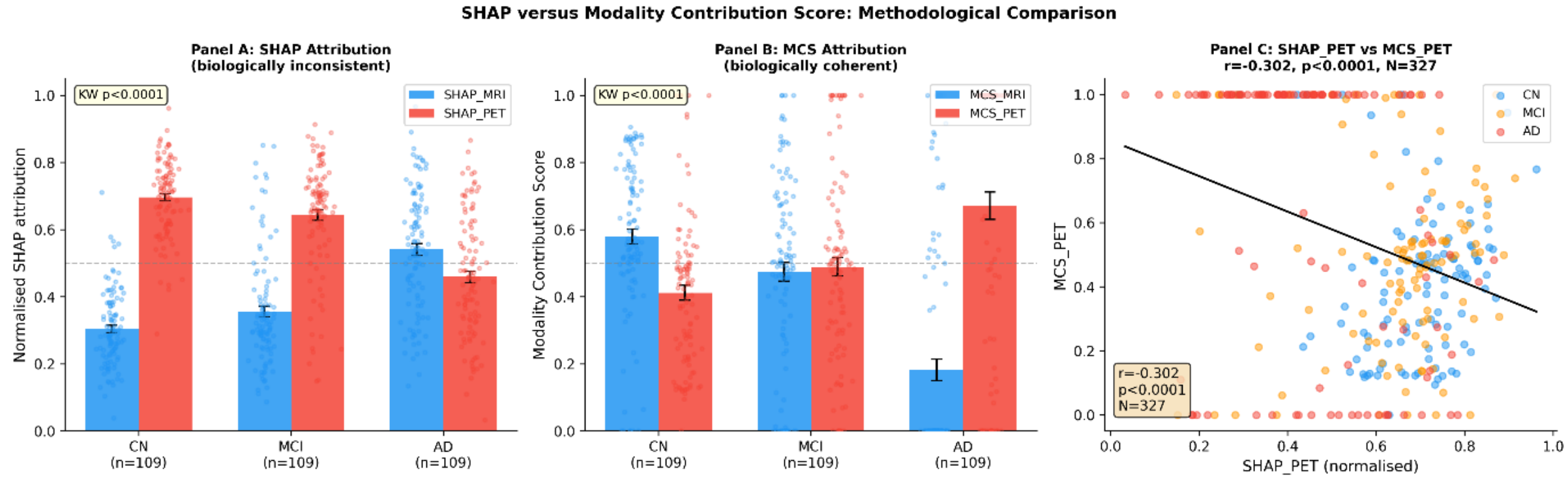

**Fig. 6. SHAP versus MCS modality attribution comparison.** Three-panel comparison of SHAP-based and ablation-based modality attribution. Panel A: Normalised SHAP modality attribution by diagnostic group showing a biologically inverted pattern with AD subjects showing higher SHAP_MRI (0.541) than CN subjects (0.304), contrary to the amyloid cascade hypothesis. Panel B: MCS modality attribution by diagnostic group showing the biologically coherent pattern with MCS_PET increasing monotonically from CN (0.412) to AD (0.671). Both panels show Kruskal-Wallis p less than 0.0001. Panel C: Scatter plot of SHAP_PET versus MCS_PET for all 327 subjects (r=minus 0.317, p less than 0.0001), demonstrating that the two methods capture fundamentally different aspects of model behaviour. Individual data points overlaid on bars represents individual subjects. CN=cognitively normal; MCI=mild cognitive impairment; AD=Alzheimer's disease; SHAP=SHapley Additive exPlanations; MCS=Modality Contribution Score.

### 3.8. OASIS-3 external validation

To assess the generalisability of the MCS framework, MCNet was applied without retraining to 1,073 OASIS-3 participants (CN = 759, AD = 314) using MRI structural features only, with PET features set to zero as shown in Figure 7. MCS_MRI was significantly higher in CN subjects ($0.756 \pm 0.148$) than in AD subjects ($0.512 \pm 0.314$; Kruskal-Wallis $H = 166.99$, $p < 0.0001$; Mann-Whitney $U$, $p < 0.0001$; $\eta^2 = 0.156$). This replicates the ADNI-3 finding that CN subjects show MRI-dominant attribution while AD subjects shift toward reduced MRI reliance, confirming that the MCS gradient is not an artefact of the ADNI training distribution but reflects a generalisable property of multimodal AD pathophysiology across independent cohorts and imaging sites. The OASIS-3 MCS_MRI values were higher overall than the ADNI-3 equivalents (CN: 0.756versus 0.579; AD: 0.512versus 0.182), which is expected for two reasons. First, OASIS-3 subjects were classified using MRI features only with PET set to zero, which structurally elevates MCS_MRI relative to the full multimodal inference used in ADNI-3. Second, OASIS-3 CN subjects are drawn from a community-based population with stringent cognitive health criteria, which may produce stronger MRI structural signals for CN classification than the ADNI-3 clinical trial enrichment protocol. Despite these distributional differences, the directional gradient (CN >AD) replicates with high statistical significance and a medium-to-large effect size ($\eta^2 = 0.156$), providing robust cross-cohort external validation of the MCS framework.

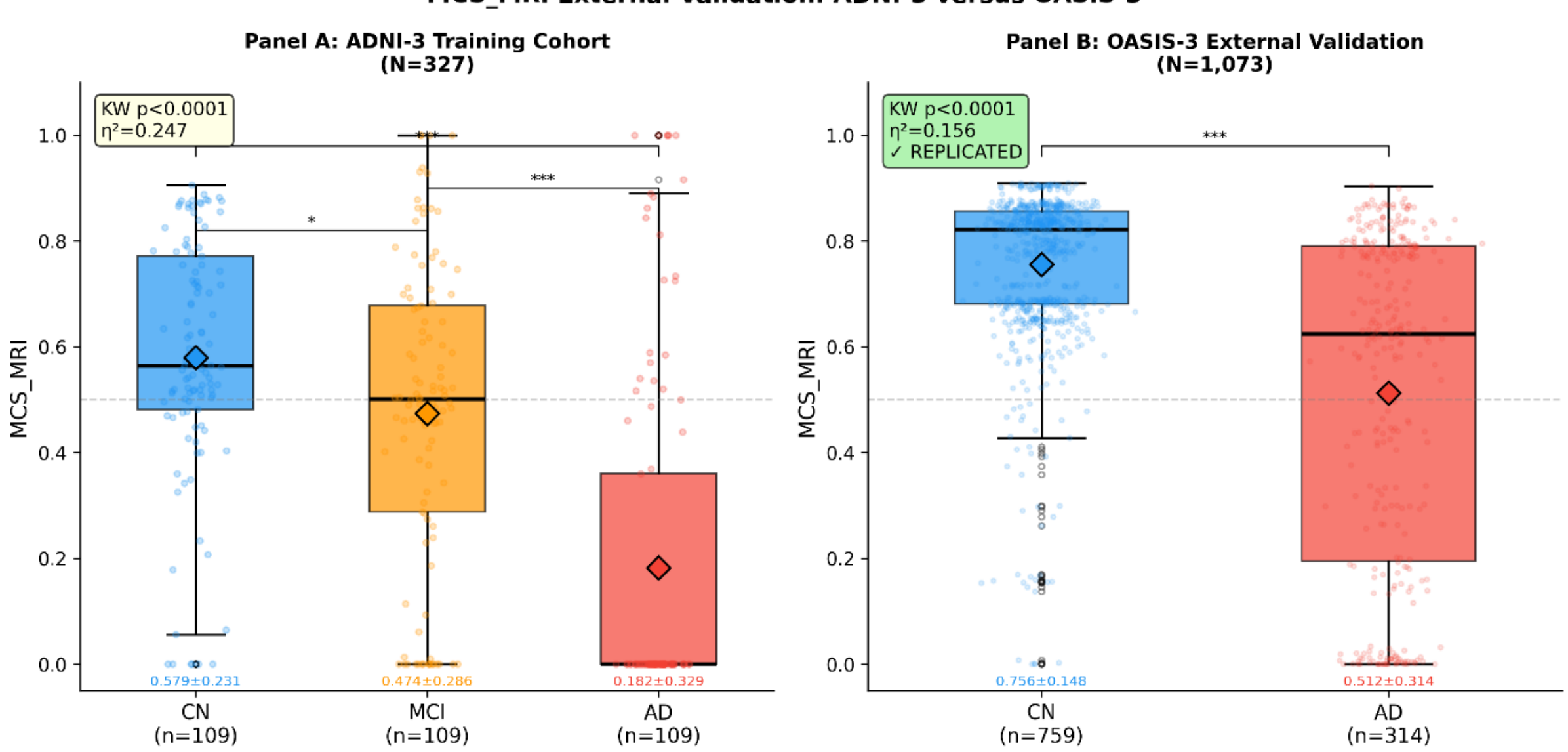

**Fig. 7. OASIS-3 external validation of MCS_MRI gradient.** Side-by-side box plots comparing MCS_MRI distributions in the ADNI-3 training cohort (Panel A, $N = 327$) and the independent OASIS-3 external validation cohort (Panel B, $N = 1{,}073$). Panel A shows the full CN, MCI, and AD gradient from the ADNI-3 cohort (Kruskal-Wallis $p < 0.0001$, $\eta^2 = 0.247$). Panel B shows the CN versus AD comparison in OASIS-3 confirming replication (CN: $0.756 \pm 0.148$versus AD: $0.512 \pm 0.314$; Kruskal-Wallis $H = 166.99$, $p < 0.0001, \eta^2 = 0.156$). Diamond markers indicate group means. Individual subject data points are overlaid. The green annotation confirms replication of the directional gradient across independent cohorts and imaging sites. CN = cognitively normal; MCI = mild cognitive impairment; AD = Alzheimer's disease; MCS = Modality Contribution Score.

## 4. Discussion

### 4.1. Classification performance

This study introduces MCNet and the MCS, a per-patient explainability framework that quantifies the relative diagnostic contribution of MRI and PET for each individual in multimodal Alzheimer's disease staging. MCNet achieved clinically competitive performance (AUC=0.881 ± 0.075, bACC=0.844 ± 0.061) for three-class CN/MCI/AD staging using pre-computed regional biomarker features from 327 balanced ADNI participants. This result is consistent with the broader literature on multimodal deep learning for AD, in which combined MRI and PET approaches consistently outperform unimodal models, particularly for the diagnostically challenging MCI category [11,12,13,15,25].The use of pre-computed UC Berkeley biomarker features rather than raw volumetric images reflects a pragmatic and reproducible methodological choice: regional biomarker summaries derived from validated processing pipelines carry well-characterised noise properties, are directly interpretable in clinical terms, and circumvent the spatial misalignment challenges that affect raw MRI-PET fusion in ADNI data, where images are acquired in scanner space on different sessions [4,36].

### 4.2. Biological coherence of the MCS gradient

The central finding of this work is that MCS_PET increased monotonically across the AD continuum (CN: 0.412 ± 0.229; MCI: 0.489 ± 0.289; AD: 0.671 ± 0.426; Kruskal-Wallis p < 0.0001), while MCS_MRI showed the inverse gradient. This pattern is biologically coherent and consistent with the amyloid cascade hypothesis, in which amyloid deposition precedes and drives tau accumulation, neurodegeneration, and cognitive impairment [34,35]. In early disease stages, structural MRI atrophy, particularly hippocampal and entorhinal volume loss, provides the primary discriminative signal for separating CN from early MCI [9,10,24]. As disease progresses, widespread amyloid deposition and cortical hypometabolism measured by PET become the dominant diagnostic signatures, consistent with the known temporal sequence of AD biomarker changes[6,9,10]. The MCS gradient observed here reflects this biological sequence at the modality level, translating a well-established pathophysiological concept into a quantitative, per-patient metric.

This is a finding of high clinical significance. A clinical diagnosis needs categorical syndromic labels but many patients do not fit into these categories and should be looked upon as unique individuals rather than simple members of this category[14]. The MCS operationalises this principle: rather than assigning a patient solely to a diagnostic category, MCNet additionally quantifies how much each imaging modality contributed to that assignment. A patient classified as MCI with MCS_MRI=0.80 has a very different imaging profile from one classified as MCI with MCS_PET=0.80, even when both receive the same diagnostic label. The heterogeneity that exists among the members of a group is pertinent and has implications for prognosis, treatment selection, and imaging follow-up strategies

The SHAP comparison provides additional mechanistic support for the MCS gradient interpretation. The inverted SHAP pattern observed in this study, in which AD subjects show higher SHAP_MRI attribution contrary to the amyloid cascade hypothesis, illustrates a fundamental limitation of reference-distribution-dependent attribution methods when applied to modality-level clinical interpretation. This finding is consistent with the broader XAI literature documenting that clinicians and developers hold opposing goals for explainability output, with clinicians requiring clinical plausibility rather than statistical deviation [37]. The MCS, by directly ablating each modality and measuring the resulting change in classification probability, produces attribution that is both statistically significant and biologically plausible, satisfying the clinical plausibility criterion that SHAP-based methods cannot meet in this setting.

### 4.3. MCI balanced attribution and diagnostic uncertainty

The MCI group demonstrated balanced attribution (MCS_MRI=0.474, MCS_PET=0.489) with neither modality clearly dominating. This finding reflects the genuine diagnostic uncertainty of MCI as a clinical construct. MCI encompasses a heterogeneous group of individuals, some progressing to AD dementia, others remaining stable or reverting to normal cognition, with variable amyloid and structural biomarker profiles [33]. The balanced MCS observed in subjects with mild cognitive impairment in this study supports the heterogeneity of the MCI with some patients with MCI being dominated by a structural signal, while in others the amyloid-metabolic signal is predominant. The group mean is, therefore, an average rather than a true biological measurement. This interpretation is supported by the representative MCI patient (Subject 012_S_6073, AV45 SUVR=1.560), who showed near-equal attribution (MCS_MRI=0.504, MCS_PET=0.496) with model confidence of only 49.6%, indicating genuine classification uncertainty rather than a confident balanced prediction.

### 4.4. Amyloid biomarker validation

The amyloid validation result (r=0.172, p=0.006, n=255) provides independent biomarker support for the MCS_PET metric, though the correlation is modest. This modesty is expected and should not be interpreted as a limitation. MCNet was trained to classify diagnostic labels (CN, MCI, AD), not to predict continuous amyloid SUVR values. MCS_PET captures the degree to which PET features drive the model's diagnostic classification, which is related to but distinct from raw amyloid burden [7,38]. The significant overall correlation confirms that subjects for whom PET features matter most in diagnosis also carry greater amyloid burden as measured by an independent pipeline[7] providing convergent validity for the MCS metric.

### 4.5. Comparison with existing explainability approaches

Existing XAI methods applied to AD neuroimaging, including SHAP, LIME, Grad-CAM, and Integrated Gradients, operate primarily at the feature or voxel level within a single modality [16,17]. They answer the question of which brain regions within an MRI or PET scan drove a classification, but they do not address which modality was more informative for a given patient. A small number of studies have explored modality weighting in multimodal fusion through attention mechanisms[13,21] but attention weights reflect learned global patterns across the training set rather than per-patient attribution derived from the actual contribution of each modality to the individual's classification probability. The MCS, computed via ablation of each modality stream (Eq. 8 to Eq. 11), directly measures the change in predicted probability when a modality is removed, providing a subject-specific, interpretable, and model-agnostic form of modality attribution not available in existing multimodal AD systems [11,12,15].

The need for multimodal XAI systems that provide clinically actionable explanations across modalities rather than within a single modality has been identified as a critical unmet need in the clinical AI literature [39]. The XAI orchestrator framework proposed by Pahud de Mortanges et al. identifies adaptivity, hierarchical explanation, and uncertainty-awareness as desirable properties of multimodal XAI systems [39]. The MCS satisfies the adaptivity criterion by computing patient-specific attribution, the hierarchical criterion by providing modality-level attribution above feature-level attribution, and contributes a quantitative uncertainty signal through the model confidence score output alongside each MCS value.

### 4.6. Clinical implications

The per-patient MCS has several potential clinical applications. In resource-constrained settings where not all imaging modalities are available, MCS trajectories from previous patients with similar profiles could guide modality prioritisation for new cases. In clinical trial design, MCS could identify imaging-phenotypic subgroups within diagnostic categories, improving trial stratification and reducing heterogeneity-driven variance in outcomes [3]. In personalised monitoring, a patient showing increasing MCS_PET over successive visits may warrant accelerated amyloid-targeted follow-up [14]. These applications require prospective validation, but the biological coherence and biomarker consistency of the MCS metric across the 327-subject ADNI cohort provide a credible foundation for such investigations.

### 4.7. Limitations

Several limitations of this study should be acknowledged. First, the cohort (n=327) is moderate in size for deep learning, and the use of pre-computed regional features rather than raw volumetric data, while methodologically sound, limits the spatial granularity of the analysis and precludes voxel-level localisation of contributing brain regions. Second, the study is cross-sectional; longitudinal validation of MCS trajectories as predictors of clinical conversion is a critical next step. Third, FDG-PET MetaROI data were available for only 143 subjects, predominantly MCI and AD, limiting the metabolic feature contribution for CN subjects; missing FDG values were imputed at the training fold mean, which may underestimate FDG contribution in CN. Fourth, the MCS is computed via linear ablation, which assumes feature independence between modalities; interaction effects between MRI and PET features may not be fully captured by this approach. Fifth, the cohort was drawn entirely from ADNI, a predominantly non-Hispanic white population; generalisability to more diverse populations require external validation [[1,3]. Sixth, the current study did not include a formal clinical validation component in which a neurologist assessed whether the MCS attribution aligns with clinical judgement for individual patients. While the statistically significant MCS gradient and convergent biomarker validations provide quantitative evidence of biological coherence, future work should incorporate blinded neurologist review of representative cases to assess whether per-patient modality attribution reflects clinical reasoning in practice. Seventh, the OASIS-3 external validation was conducted using MRI structural features only, as pre-computed PET SUVR data were not available for the full cohort in a format compatible with the MCNet pipeline. Raw PUP processed PET files were impractical to download at approximately 5.3 GB per subject. The replication of the MCS_MRI gradient is therefore partial; full multimodal external validation including PET features in an independent cohort remains an important next step.

### 4.8. Future directions

Longitudinal validation is the most pressing next step, tracking MCS trajectories in ADNI participants who subsequently convert from CN to MCI or MCI to AD to determine whether rising MCS_PET precedes clinical conversion and whether the rate of MCS shift correlates with conversion speed. Full multimodal external validation incorporating PET SUVR features in an independent cohort with pre-computed amyloid and metabolic data remains an important priority, as the current OASIS-3 validation was necessarily restricted to MRI features only. Integration of tau PET as a fourth feature stream, leveraging the 249 ADNI-3 subjects with available UC Berkeley AV1451 SUVR data identified in this study, would further align the MCNet framework with the current NIA-AA biological staging criteria for AD and extend the MCS to a three-modality attribution (MCS_MRI, MCS_amyloid, MCS_tau) [23]. Extension to more demographically diverse cohorts beyond the predominantly non-Hispanic white ADNI population is necessary to establish generalisability across ancestry groups, consistent with Lancet Commission recommendations on dementia prevention and care [3]. Finally, prospective neurologist validation of MCS-guided imaging recommendations in a clinical decision support setting would provide the implementation evidence needed for real-world healthcare adoption, addressing the gap between statistical validation and clinical actionability identified in the XAI literature [37,39].

## 5. Conclusion

This study introduces per-patient modality attribution as a technically novel and clinically interpretable dimension of explainability in multimodal Alzheimer's disease staging, addressing a fundamental limitation that no existing multimodal AI system has resolved. Rather than producing a classification label alone, MCNet additionally quantifies for each individual patient the degree to which structural brain atrophy on MRI versus amyloid burden and metabolic dysfunction on PET drove the diagnostic decision. This shift from population-level to patient-level explainability is clinically meaningful: two patients sharing the same diagnostic label may have fundamentally different imaging profiles, and the MCS makes this heterogeneity visible, interpretable, and actionable. The biologically coherent MCS gradient across the CN-MCI-AD continuum precisely mirrors the known temporal sequence of AD pathophysiology, reflecting the progressive dominance shift from structural neurodegeneration to amyloid and metabolic PET signal as disease advances. Critically, this gradient is not a training artefact but replicates without retraining in an independent external cohort of over one thousand subjects, demonstrating genuine cross-cohort generalisability that strengthens confidence in the biological validity of the framework. The mechanistic comparison with SHAP further establishes that reference-distribution-dependent attribution methods are fundamentally unsuitable for modality-level clinical interpretation, while ablation-based MCS satisfies the clinical plausibility criterion that clinicians require from AI-assisted diagnostic tools. Looking forward, longitudinal validation of MCS trajectories as early predictors of clinical conversion, integration of tau PET as a third modality stream aligned with current NIA-AA biological staging criteria, and prospective neurologist validation of MCS-guided imaging recommendations represent the most pressing next steps toward clinical translation. Taken together, the present findings position the MCS framework as a biologically grounded, statistically robust, and externally validated foundation for trustworthy, personalised, and clinically actionable AI in multimodal AD neuroimaging, one that advances the field not by improving classification accuracy alone, but by making each diagnostic decision interpretable at the individual patient level.

## Acknowledgments

Data collection and sharing for this project was funded by the Alzheimer's Disease Neuroimaging Initiative (ADNI) (National Institutes of Health Grant U01 AG024904) and DOD ADNI (Department of Defense award number W81XWH-12-2-0012). Further information about ADNI can be found at adni.loni.usc.edu. ADNI is funded by the National Institute on Aging, the National Institute of Biomedical Imaging and Bioengineering, and through generous contributions from AbbVie, Alzheimer's Association, Alzheimer's Drug Discovery Foundation, Araclon Biotech, BioClinica Inc., Biogen,

Bristol-Myers Squibb Company, CereSpir Inc., Cognitect, Eisai Inc., Elan Pharmaceuticals Inc., Eli Lilly and Company, EuroImmun, F. Hoffmann-La Roche Ltd and its affiliated company Genentech Inc., Fujirebio, GE Healthcare, IXICO Ltd., Janssen Alzheimer Immunotherapy Research and Development LLC., Johnson and Johnson Pharmaceutical Research and Development LLC., Lumosity, Lundbeck, Merck and Co. Inc., Meso Scale Diagnostics LLC., NeuroRx Research, Neurotrack Technologies, Novartis Pharmaceuticals Corporation, Pfizer Inc., Piramal Imaging, Servier, Takeda Pharmaceutical Company, and Transition Therapeutics. The Canadian Institutes of Health Research provides funds to support ADNI clinical sites in Canada. Private sector contributions are facilitated by the Foundation for the National Institutes of Health (www.fnih.org). The grantee organisation is the Northern California Institute for Research and Education, and the study is coordinated by the Alzheimer's Therapeutic Research Institute at the University of Southern California. ADNI data are disseminated by the Laboratory of Neuro Imaging at the University of Southern California. OASIS-3 data were provided by the Open Access Series of Imaging Studies. OASIS-3 is supported by grants P30 AG066444, P50 AG00561, P30 NS09857781, P01 AG026276, P01 AG003991, R01 AG043434, UL1 TR000448, and R01 EB009352. The authors thank all ADNI and OASIS-3 participants and their families for their invaluable contribution to research.

**Consent statement**

This study used existing de-identified data from the Alzheimer's Disease Neuroimaging Initiative (ADNI) database and the Open Access Series of Imaging Studies (OASIS-3) repository. All ADNI and OASIS-3 participants provided written informed consent at the time of original data collection at their respective participating sites. The ADNI study was approved by the institutional review boards of all participating sites and conducted in accordance with the Declaration of Helsinki. The OASIS-3 study was approved by the Washington University Institutional Review Board and conducted in accordance with ethical guidelines at all participating institutions. OASIS-3 data were accessed under an approved data use agreement in compliance with the OASIS data sharing policy. No new participant recruitment or data collection was performed for this study.